\documentclass[fleqn,10pt]{wlscirep}
\usepackage[utf8]{inputenc}
\usepackage[T1]{fontenc}
\usepackage{graphicx}
\usepackage{amssymb}
\usepackage{graphicx}
\usepackage{booktabs}
\usepackage{listings}
\usepackage{xcolor}
\usepackage{hyperref}
\usepackage{caption}
\usepackage{subcaption} 
\usepackage{url}
\usepackage{color}
\usepackage{subfiles}
\usepackage{cite}
\title{Generative Adversarial Networks based Skin Lesion Segmentation}

\author[1,*]{Shubham Innani}
\author[1]{Prasad Dutande}
\author[1,2]{Ujjwal Baid}
\author[3]{Venu Pokuri}
\author[2]{Spyridon Bakas}
\author[1]{Sanjay Talbar}
\author[1,2,$\dagger$]{Bhakti Baheti}
\author[3,$\dagger$]{Sharath Chandra Guntuku}

\affil[1]{Center of Excellence in Signal and Image Processing, Shri Guru Gobind Singhji Institute of Engineering and Technology, Nanded, Maharashtra, India}
\affil[2]{Center for Biomedical Image Computing and Analytics, University of Pennsylvania, Philadelphia, United States}
\affil[3]{Department of Computer and Information Science,  University of Pennsylvania, Philadelphia, United States}
\affil[*]{corresponding author (shubham.innani@gmail.com)}
\affil[$\dagger$]{Co-supervising authors}

\begin{abstract}
Skin cancer is a serious condition that requires accurate diagnosis and treatment. One way to assist clinicians in this task is using computer-aided diagnosis (CAD) tools that automatically segment skin lesions from dermoscopic images. We propose a novel adversarial learning-based framework called Efficient-GAN (EGAN) that uses an unsupervised generative network to generate accurate lesion masks. It consists of a generator module with a top-down squeeze excitation-based compound scaled path, an asymmetric lateral connection-based bottom-up path, and a discriminator module that distinguishes between original and synthetic masks.
A morphology-based smoothing loss is also implemented to encourage the network to create smooth semantic boundaries of lesions. The framework is evaluated on the International Skin Imaging Collaboration (ISIC) Lesion Dataset 2018. It outperforms the current state-of-the-art skin lesion segmentation approaches with a Dice coefficient, Jaccard similarity, and Accuracy of 90.1\%, 83.6\%, and 94.5\%, respectively. We also design a lightweight segmentation framework (MGAN) that achieves comparable performance as EGAN but with an order of magnitude lower number of training parameters, thus resulting in faster inference times for low compute resource settings. 
\end{abstract}
\begin{document}

\flushbottom
\maketitle 
%
%
\thispagestyle{empty}


\section*{Introduction}

Skin cancer results in approximately 91,000 deaths annually \cite{wiki2}. Early detection and regular monitoring are crucial in improving the quality of diagnosis, ensuring accurate treatment planning, and reducing skin cancer mortality rates~\cite{Rigel}. A common detection method involves a dermatologist examining skin images to identify ambiguous clinical patterns of lesions that are often not visible to the naked eye. Dermoscopy, a widely used technique, helps dermatologists differentiate between malignant and benign lesions by eliminating surface reflections on the skin, thereby improving the accuracy of skin cancer diagnosis \cite{wiki3}.

\begin{figure}[b]
	\begin{center}
		\captionsetup{justification=centering}
		\includegraphics[scale=0.35]{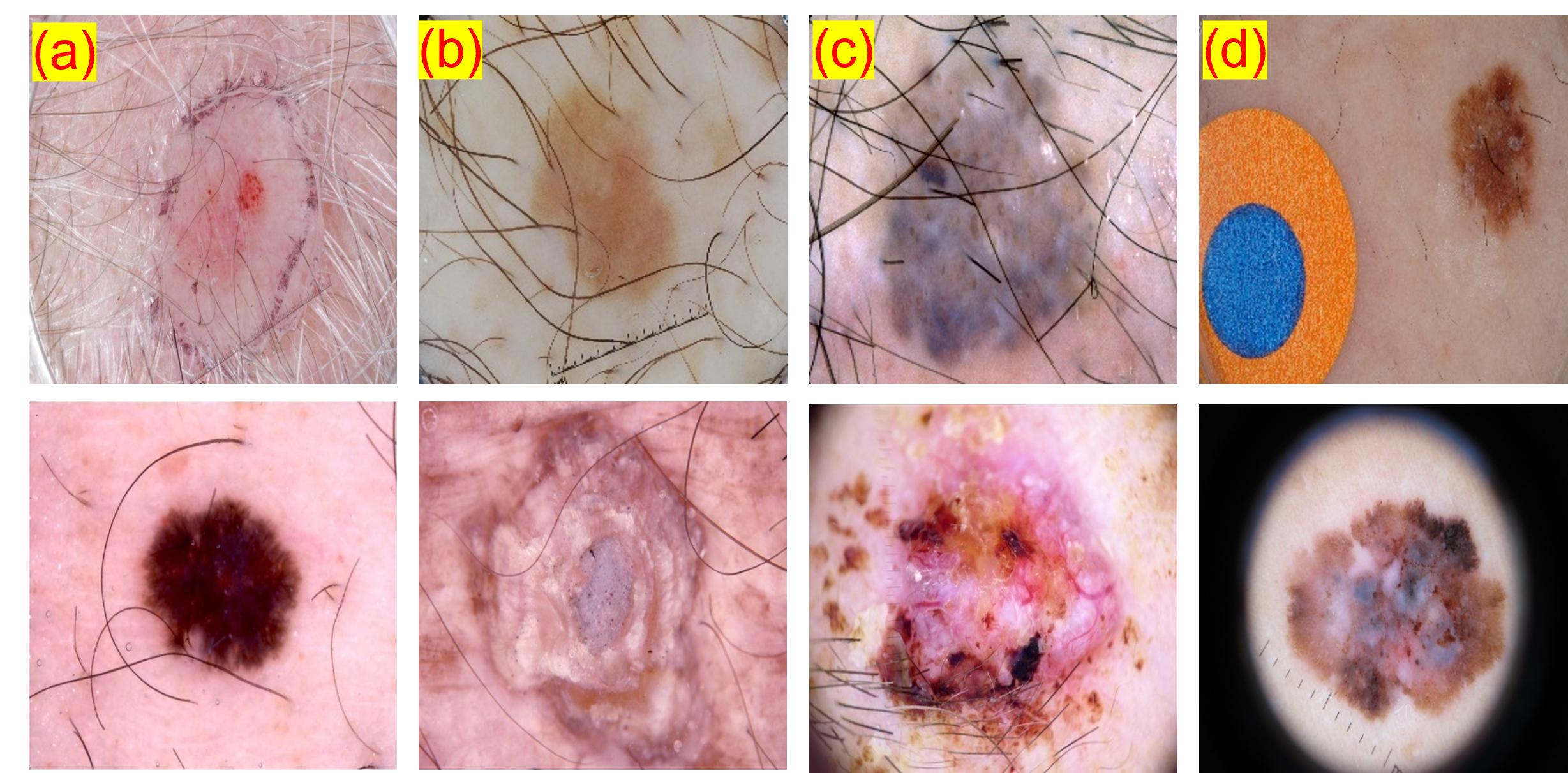}
		\caption{Challenges in skin lesion segmentation using dermoscopic images. First row: a) minor variation in the lesion and skin color, b) low contrast between wound and skin, c) occlusion in lesions due to hair, and d) artifacts from image acquisition. Second row: a few examples from the ISIC Lesion dataset ~\cite{ISIC2018} used in this paper.}
		\label{database_img}
	\end{center}
\end{figure}

\begin{table*}[t]
\centering
\caption{Related work on skin lesion segmentation with CNN and GAN-based approaches}
\label{related_work}
\scalebox{0.8}{
\begin{tabular}{|r|l|c|} 
\hline
\textbf{Model name / Citation}                                                                               & \textbf{One phrase description}                                                       & \textbf{Architecture}  \\ 
\hline
Saliency Maps \cite{Jahanifar}                                                  & Segmentation based on Supervised Saliency Maps                         & Classical       \\ 
\hline
UNet Segmentation \cite{TANG2019289}                                            & Stochastic weight averaging using UNet                                 & CNN             \\ 
\hline
Deep CNN \cite{ALMASNI2018221}                                                  & Full Resolution Networks                                               & CNN             \\ 
\hline
BLA-Net \cite{FENG2022107190}               & deformable convolution ResNet34 with auxiliary boundary learning network                               & CNN             \\ 
\hline
ERU\cite{9303084}                                                              & EfficientNetB4 with UNet based encoder-decoder                         & CNN             \\ 
\hline
AS-Net\cite{HU2022117112}                                                      & Combines spatial and channel attention for learning                    & CNN             \\ 
\hline
Attention Network \cite{XIE2020105241}                                          & Attention mechanism with high resolution features                       & CNN             \\ 
\hline
SEACU-Net \cite{JIANG2022107076}                                                & Squeeze and Excitation based Attentive ConvLSTM                       & CNN             \\ 
\hline
Conditional Random Fields \cite{sr_skin}                                                & Deep Learning Approach with Pre and Post Processing                       & CNN             \\ 
\hline  
FAT-Net \cite{WU2022102327}                                                     & Feature Adaptive Transformers                                          & Transformer     \\ 
\hline
DFE-Net \cite{FAN2023104423}                                                    & CNN and Transformer based Feature extraction                           & Transformer     \\ 
\hline
SLT-Net \cite{FENG2022105942}                                                   & CSwin Transformer replaced Conv module in UNet                         & Transformer     \\ 
\hline
cGAN \cite{cgan}                                                                & Conditional Generative Adversarial Network                             & GAN             \\ 
\hline
Generative Network \cite{8363712}                                               & Decisive Generator for skin lesion segmentation                        & GAN            \\ 
\hline
DCGAN \cite{DCGAN}                                                              & Generating Synthetic Skin Images                                       & GAN             \\ 
\hline
FCA-Net \cite{fcanet}                                                           & Factorised channel attention and multi-scale features                  & GAN             \\ 
\hline
DAGAN \cite{LEI2020101716}                                                      & Deep Neural Network with generative Networks                           & GAN             \\ 
\hline
UNet-SCDC GAN \cite{LEI2020101716}                                              & Leveraging power of discriminators                                       & GAN             \\ 
\hline
SLS-Net \cite{SARKER2021115433}                                                 & Lightweight device model with GAN                                      & GAN             \\
\hline
\end{tabular}}
\end{table*}

\noindent Skin lesion segmentation, a method to differentiate foreground lesions from the background, has received a lot of attention for over a decade due to its high clinical applicability and demanding nature. Computer-aided diagnostic algorithms for automated skin lesion segmentation could aid clinicians in precise treatment and diagnosis, strategic planning, and cost reduction. However, automated skin lesion segmentation is challenging due to several factors \cite{ALMASNI2018221} such as (i) large variance in shape, texture, color, geographical conditions, and fuzzy boundaries, (ii) the presence of artifacts such as hair and blood vessels, and (iii) poor contrast between background skin and cancer lesions in addition to artifacts from image acquisition, as shown in Figure \ref{database_img}.

\subsection*{Prior Work} 
Pixel-level skin lesion segmentation algorithms can be divided into approaches built upon a) classical image processing and b) deep learning-based architectures. Deep learning-based methods can be further classified into Convolutional Neural Networks (CNN) and Adversarial Learning-based Generative Networks (GAN) based on the network topology. A brief review of a few prior works in these categories is presented in Table \ref{related_work}. 
The performance of classical image processing approaches heavily depends on post-processing, such as thresholding, clustering, and hole filling, tuning hyperparameters, and manual feature selection. Manually tuning these parameters can be expensive and could result in poor generalizability. Lately, deep learning-based approaches have surpassed several classical image processing-based approaches, mainly due to the wide availability of large labeled datasets and compute resources.
Deep convolutional neural networks (DCNN) based methods gained a lot of popularity for skin lesion segmentation prior to the introduction of Transformer and GAN-based approaches in the field of medical imaging,~\cite{yi2019generative,lihan,yao2020comprehensive,LIU2023106874,WU2023106457}. 

\begin{figure*}[!b]
	\begin{center}
		\captionsetup{justification=centering}		\includegraphics[scale=0.5]{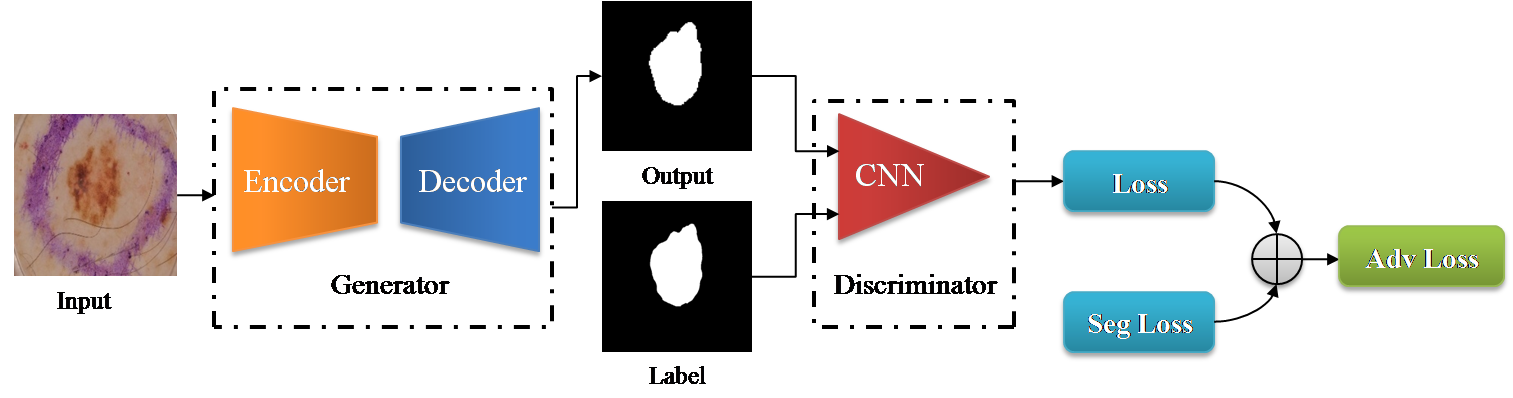}
		\caption{Flowchart of the proposed framework. The generator module is an encoder-decoder network. The discriminator classifies the segmentation result as real or fake.}
		\label{architecture_gan}
	\end{center}
\end{figure*}

\noindent The success of prior DCNN-based approaches in skin lesion segmentation is primarily based on supervised methods that rely on large labeled datasets to extract features related to the image's spatial characteristics and deep semantic maps. However, gathering a large dataset with finely annotated images is time-consuming and expensive.
To address this challenge, Goodfellow \textit{et al.} \cite{goodfellow} introduced Generative Adversarial Networks (GANs), which have gained popularity in various applications, including medical image synthesis, due to the lack of widely available finely annotated data. Several recent and relevant GAN-based approaches in skin lesion analysis from the literature are listed in Table \ref{related_work}. Unsupervised learning-based algorithms that can handle large datasets with precision and high performance without requiring ground truth labels carry significant promise in addressing real-world problems such as computer-aided medical image analysis. 


\noindent In our work, we address the challenges of lesion segmentation by utilizing generative adversarial networks (GANs) \cite{goodfellow}, which can generate accurate segmentation masks with minimal or no supervision. GANs work by training a generator and discriminator to compete against each other, where the generator tries to create realistic images, and the discriminator tries to differentiate between real and generated images (Figure ~\ref{architecture_gan}). However, designing an effective GAN for segmentation takes considerable time, as the performance is highly dependent on the architecture and choice of the loss function. Our study aims to optimize all three components (generator, discriminator, and loss function) for better segmentation results. The choice of the loss function is critical for the success of any deep learning architecture, and our approach takes this into account\cite{pmlr-v102-kervadec19a}.

\subsection*{Proposed Work} 
We propose two GAN frameworks for skin lesion segmentation. The first is {\color{red}Efficient-GAN (EGAN)}, which focuses on precision and learns in an unsupervised manner, making it data-efficient. It uses an encoder-decoder-based generator, patchGAN \cite{patchgan} based discriminator and smoothing-based loss function. The generator architecture uses a squeeze and excitation-based compound scaled encoder and a lateral connection-based asymmetric decoder. This architecture captures dense features to generate fine-grained segmentation maps, and the discriminator distinguishes between synthetic and original labels. We also implement a morphological-based smoothing loss function to capture fuzzy boundaries more effectively.

\noindent Although deep learning methods provide high precision for lesion segmentation, they are computationally expensive, making them impractical for real-world applications with limited resources like dermatoscopy machines. This presents a challenge in contexts where high-resource devices are unavailable to dermatologists. To address this issue, various devices like MoleScope II, DermLite, and HandyScope have been developed for lesion analysis and support low computational resources. These devices use a special lens with a smartphone. To create a more practical model for such real-time applications, we propose {\color{red}Mobile-GAN (MGAN)}, which is a lightweight unsupervised model consisting of an Inverted Residual block\cite{MobileNetV2}  with Atrous Spatial Pyramid Pooling\cite{Deeplabv3}. This model aims to achieve good segmentation performance in terms of the Jaccard score with lower resource strain. With only 2.2M parameters (as opposed to 27M parameters in EGAN), the model can run at 13 frames per second, increasing the potential impact of computer vision-based approaches in day-to-day clinical practice.






\section*{Results} \label{results}

\subsection*{Performance of CNN-based models}
 We implemented and analyzed the results of several CNN and GAN-based approaches for this task. Table \ref{cnn_results} summarizes the evaluation of  CNN and GAN-based approaches on the unseen test dataset. We started with one of the most popular architectures in medical imaging segmentation - UNet ~\cite{Ronneberger}. Since this architecture is a simple stack of convolutional layers, the original UNet provided a baseline performance on ISIC 2018 dataset. We strategically conducted several experiments using deeper encoders like ResNet, MobileNet, EfficientNet, and asymmetric decoders (described in the Methods section)
The concatenation of low-level features is skewed rather than linking each block from the encoder, like in traditional UNet. Adding a batch normalization layer after each convolutional layer also helped achieve better performance. For detailed evaluation with CNN-based methods, we also experiment with DeepLabV3+ \cite{Deeplabv3} and Feature Pyramid Network (FPN) \cite{fpn} decoders in combination with various encoders as described above, and the modification led to improved performance. These results on the ISIC 2018 test set from our experimentation, i.e., us running the authors' code to train the proposed models, are listed with $*$  in Table \ref{cnn_results}. 
 
\subsection*{Performance of GAN-based models}
 Table  \ref{cnn_results} also lists several results from recent literature on this dataset for comparison completeness. Models trained by us are submitted to the evaluation server for a fair evaluation. We then compare the results of various GAN-based approaches, as shown in Table \ref{cnn_results}. 
 We observe that a well-designed generative adversarial network (GAN) improves performance compared to techniques based on CNNs for medical image segmentation. This is because of GANs ability to overcome the main challenge in this domain of not having large labeled training data. Our proposed EGAN approach outperforms all other approaches. 
 A few works\cite{FENG2022107190,HAN2023106343,cpfnet,9303084} report better performance compared to our results. But these works created and used an independent test split from ISIC training data and did not use the actual ISIC test data. 

\begin{table*}[t]
\centering
	
	\caption{Results of CNN and GAN-based approaches including our proposed algorithms (MGAN and EGAN) on the ISIC 2018 test dataset. * indicates the model was re-trained using the authors' source code. - indicates metrics not being reported}
	\label{cnn_results}
	\scalebox{0.75}{
	\begin{tabular}{ccccccc}
\hline
\textbf{Approach}                       & \textbf{Network} & \textbf{Dice Coefficient} & \textbf{Accuracy} & \textbf{Jaccard Index} & \textbf{Sensitivity} & \textbf{Specificity} \\ \hline
UNet*  \cite{Ronneberger}               & CNN               & 71.53                       & 80.7              & 60.58                   & 85.8                 & 87.8                 \\
DeepLabV3+* \cite{prl}                 & CNN               & 76.3                      & 91.4              & 77.3                     & 87.8                 & 88.1                 \\
FPN* \cite{agri}                       & CNN               & 84.46                      & 92.4              & 73.76                 & 88.2                 & 87.8                 \\
Mobile-UNet* \cite{innani2023deep}            & CNN               & 83.9                      & 91.2              & 71.32                   & 90                   & 94.1                 \\
Res-UNet* \cite{innani2023deep}            & CNN               & 86.3                      & 90.8              & 76.77                   & 90.9                 & 93.7                 \\
Eff-UNet*  \cite{effunet}             & CNN               & 89.56                      & 92.2              & 81.42                   & 90.7                 & 92.6               \\
CPFNet  \cite{cpfnet}           & CNN               & 89.89                     & 96.3             & 82.86                  & 89.53                & 96.55                \\
ERU  \cite{9303084}              & CNN               & 88.12                     & 94.35             &   80.56                & 90.32                & 96.92                \\
FAT-Net \cite{WU2022102327}           & Transformers      & 89.03                     & 96.99             & 82                     & 91                   & 95.3                 \\
SEACU-Net \cite{JIANG2022107076} & CNN                         & 87.58                       & 93.60                       & 78.12                       & -                            & -                            \\
SLT-Net \cite{FENG2022105942}    & Transformer                 & 82.85 & -    &  71.51 & -                            & -                            \\
AS-Net \cite{HU2022117112}       & CNN                         & 88.07                       & 94.66                       & 80.51                       & 89.92                       &                       95.72 \\
\hline
cGAN \cite{LEI2020101716}    & GAN               & 83.8                      & 90.2              & 74.8                   & 89.2                 & 92.9                 \\
DCGAN \cite{DCGAN}           & GAN               & 85.8                      & 91.9              & 74.4                   & 89.1                 & 95.4                 \\
FCA-Net \cite{fcanet}        & GAN               & 88.8                      & 93.8              & 77.2                   & 94.7                 & 92.1                 \\
DAGAN \cite{LEI2020101716}           & GAN               & 88.5                      & 92.9              & 82.5                   & \textbf{95.3}                 & 91.1                 \\
\textbf{MGAN (Ours)}     & GAN     & 88.3             & 93.4     & 75.0         & 93.8        & 92.1      \\
\textbf{EGAN (Ours)}  & GAN      & \textbf{90.1}             & \textbf{94.5}     & \textbf{83.6}          & 93.6        & \textbf{95.5}       \\ \hline
\end{tabular}}
\end{table*}

\subsection*{Performance of lightweight models}
We designed a lightweight generator model called MGAN, based on DeepLabV3+ and MobileNetV2, which achieves results comparable to our EGAN model in terms of Dice Coefficient with significantly fewer parameters and faster inference times. Table \ref{mobile_inference} comprehensively compares various mobile architectures based on the Jaccard Index, the number of parameters in a million, and inference speed on the test dataset for a patch size of $512 \times 512$. As we can see from the table \ref{mobile_inference} MGAN has 2.2M parameters providing the Inference Speed of 13FPS.
Even though SLSNet reports a higher performance in terms of the Jaccard Index, this result is evaluated on the independent validation test set. 

\begin{table}[h]
\centering
	
	\caption{Comparison of various Mobile networks available in the literature. The inference column indicates the Frames per Second (FPS) on original Images with a patch size of $512 \times 512$}
	\label{mobile_inference}
	\scalebox{0.85}{
\begin{tabular}{cccc}

\hline
Method             & Jaccard Index & Parameters (M)   & Inference (FPS) \\ \hline
ENet\cite{enet}              & 72.7          & 0.36         & 27              \\
SLSNet\cite{SARKER2021115433}             & 78.40         & 2.35         & 16              \\
GAN FCN\cite{GANFCN}           & 77.80         & 10.2         & 8               \\
\textbf{MGAN} & \textbf{75} & \textbf{2.2} & \textbf{13}     \\ \hline
\end{tabular}}
\end{table}

\begin{figure}[h]
	\begin{center}
		\includegraphics[scale=0.5]{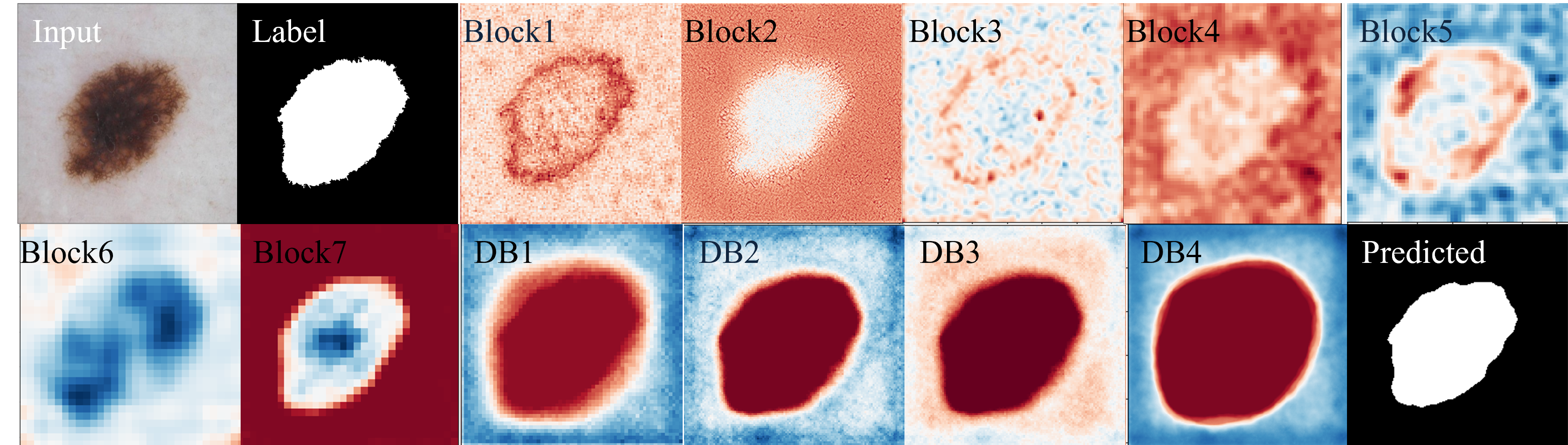}
		\caption{Visualization of the feature maps of proposed EGAN architecture. 
  }
		\label{feature_maps}
	\end{center}
\end{figure}

\begin{figure}[t]
	\begin{center}
		\includegraphics[scale=0.5]{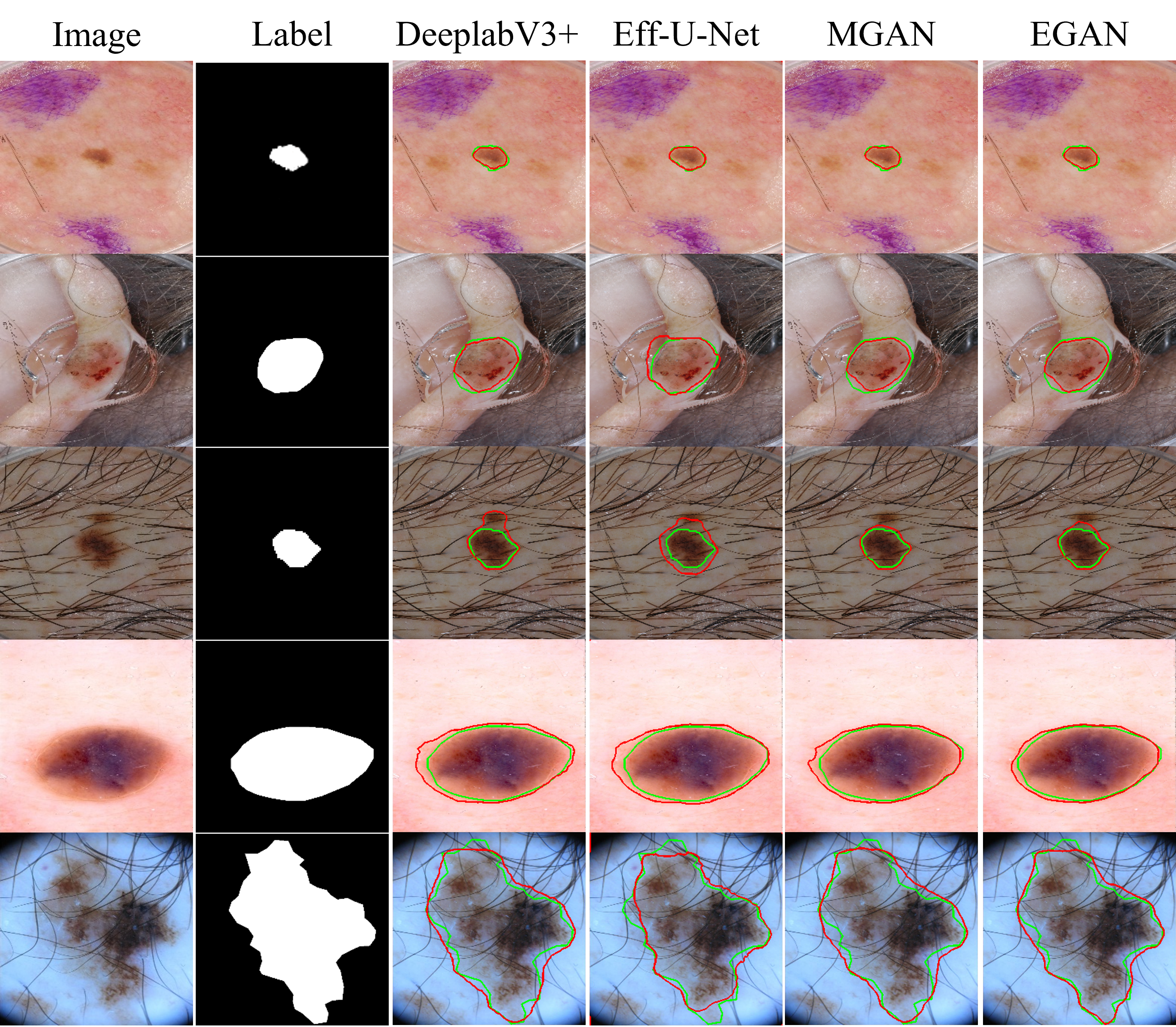}
		\caption{Comparison of the segmentation by various CNN and GAN-based approaches. Each column serially depicts the input image, label, output of various CNN-based approaches, and output of proposed MGAN and EGAN. 
 Ground truth and segmented lesions are marked with green and red curves respectively.}
		\label{seg_op1}
	\end{center}
\end{figure}

\subsection*{Visualization of the learned representations}

One of the criticisms of deep neural networks, which can make valuable and skillful predictions, is that they are generally opaque, i.e., it is unclear how or why a particular prediction or decision is made. To address concerns about the opacity of deep neural networks, we utilized the internal structures of convolutional neural networks that work on 2D image data to investigate the representations learned by our unsupervised model. Figure \ref{seg_op1} displays the segmentation results for visual interpretation. The proposed GAN framework also demonstrates better segmentation performance regardless of non-skin objects or artifacts in the image. We assessed and visualized the 2D filter weights of the model to explore the features learned by the model. Additionally, we investigated the activation layers of the model to understand precisely which features the model recognized for a given input image, and we displayed the results in Figure \ref{feature_maps} for visual understanding. We selected the output of seven blocks of the encoder (Block1-Block7) and four output feature maps from the decoder (D1-D4) for visualization, as the model has numerous convolutional layers in each architecture block.

\section*{Discussion} \label{discussion}

This paper has three main findings. First, we proposed a novel unsupervised adversarial learning-based framework (EGAN) based on Generative Adversarial Networks(GANs) to segment skin lesions in a fine-grained manner accurately. In data-scarce applications such as skin lesion segmentation, the success of GANs relies on the quality of the generator, discriminator, and loss function used. 
One of the main challenges in the field of medical imaging is the availability of large annotated data, collecting which is a tedious, consuming, and costly task. To address the data-efficiency challenge, we trained our model unsupervised, allowing the generator module to capture features effectively and segment the lesion without supervision. Our patchGAN-based discriminator penalized the adversarial network by differentiating between labels and predictions. As we do not backpropagate the error during training in the discriminator, no such advancement is needed as PatchGAN-based architecture is powerful enough to classify between real and fake. In skin lesion segmentation, capturing contextual information around the segmentation boundary is crucial for improving performance \cite{FENG2022107190}. To address this, we also implemented the morphological-based smoothing loss to capture fuzzy lesion boundaries, resulting in a highly discriminative GAN that considers contextual information and segmented boundaries. 
The performance-exclusive EGAN approach outperforms prior works achieving improved performance with a dice coefficient of 90.1\% on the ISIC 2018 test dataset when trained with adversarial learning and morphology-based smoothing loss function compared to using the dice loss alone, which achieved a dice coefficient of 88.4\% revealing the potential of our methodology. Our evaluation of the ISIC 2018 dataset demonstrates significantly improved performance compared to existing models in the literature. Furthermore, the proposed framework's potential can be extended to other medical imaging applications. 

\noindent Second, we proposed a lightweight segmentation framework (MGAN) that achieves comparable results while being much less computationally expensive -- with an order of magnitude lower number of training parameters and significantly faster inference time. 
The MGAN approach is suitable for real-time applications, making it a viable solution for cutting-edge deployment, for instance, in low compute resource contexts. Our proposed framework includes two generative models: EGAN and MGAN, which are designed to balance performance and efficiency. Integrating models like MGAN with dermoscopy devices can revolutionize the future of dermatology, enabling more efficient, accurate, real-time segmentation and accessible care for patients with skin lesions.

\noindent Third, our approach enables visualizing the learned representations of the model to interpret the predictions. This is especially crucial for \textit{clinical algorithms-in-the-loop} applications such as skin lesion segmentation, where the decisions of automated segmentation methods could be considered by clinicians in the context of the features learned by the model.   


\paragraph{Limitations:} Although our model has achieved promising performance on ISIC 2018 dataset, the performance could not be evaluated on other datasets. We explored different datasets such as Derm7pt\cite{8333693}, Diverse Dermatology Images\cite{ddt}, and Fitzpatrick 17k\cite{Fitzpatrick}, among others, to assess the generalizability of the proposed approach. However, we noticed that segmentation masks were not available. While segmentation masks were available for the PH2 dataset\cite{6610779},  we could not access the dataset. Deep Learning models are computationally intensive and require significant resources. EGAN model is computationally heavy for deployment in real-time clinical applications. This can limit the use in resource constraint environments or devices with limited processing capabilities. In such scenarios, models such as MGAN could be utilized.  

\section*{Methods} \label{ref:methodology}

The skin lesion GAN-based segmentation framework we propose in this work is shown in Figure \ref{architecture_gan}. The framework contains three main components: i) the generator, which consists of an encoder to extract feature maps and a decoder to generate segmentation maps without supervision and adapt to variations in contrast and artifacts; ii) the discriminator, which distinguishes between the reference label and the segmentation output; and iii) appropriate loss functions to prevent overfitting, achieve excellent convergence, and accurately capture fuzzy lesion boundaries.

\subsection*{Dataset}

The proposed segmentation approach was evaluated using the ISIC 2018 dataset, a standard skin lesion analysis dataset. This dataset contains 2594 images with corresponding ground truth, of which 20\% (approximately 514 images) were used for validation. The images in the dataset vary in size and aspect ratio and contain lesions with different appearances in various skin areas. Some sample images from the dataset are shown in Figure \ref{database_img}. To ensure a fair evaluation, the results of the test set were uploaded to the online server of the ISIC 2018\cite{ISIC2018} portal. 

\begin{figure}[!t]
    \centering
     \begin{subfigure}{1\textwidth}
     \centering
         {\includegraphics[scale=0.5]{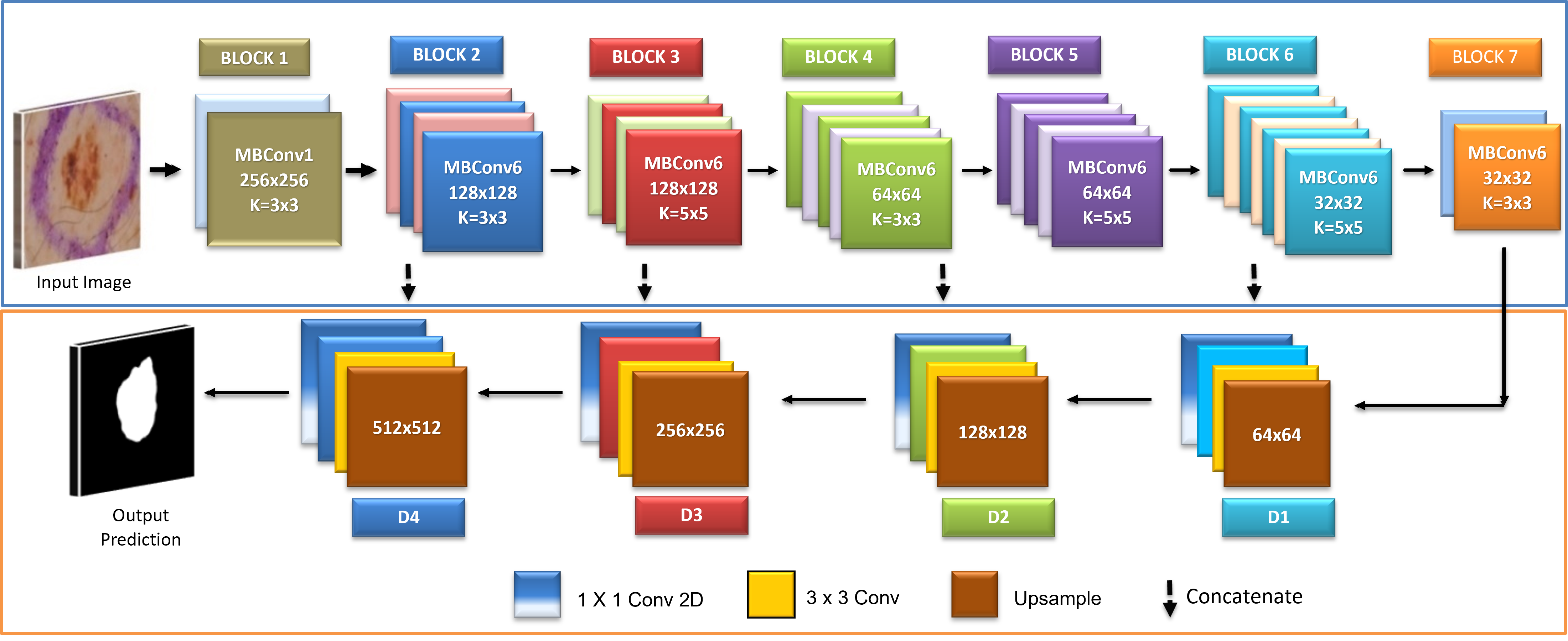}}
         \caption{Encoder has seven blocks and asymmetric concatenation to the decoder with lateral connections. Decoders made of four blocks D1-D4 of Conv2D.}
         \label{block_diagram}
     \end{subfigure}

     \begin{subfigure}{0.45\textwidth}
     \centering
         {\includegraphics[scale=0.4]{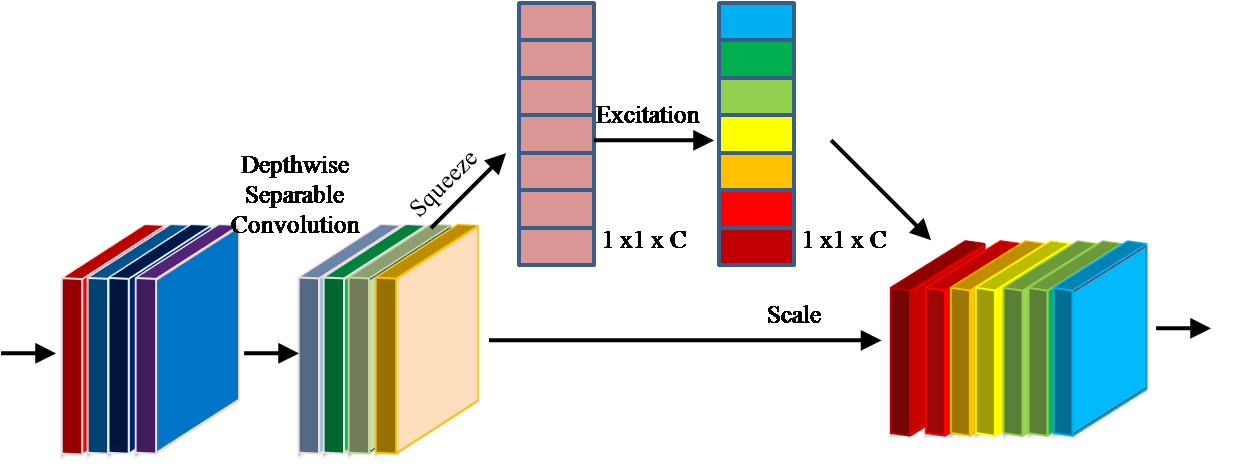}}
         \caption{MBConv Block of the Encoder}
         \label{mbconv}
     \end{subfigure}
     \caption{The architecture of the proposed generator in the EGAN architecture.}
     \label{seg}
\end{figure} 

\subsection*{Generative Adversarial Network}

Goodfellow \textit{et al.} \cite{goodfellow} first introduced Generative adversarial networks (GAN) to generate synthetic data. Labeling clinical information is a tricky and time-consuming task requiring a specialist. Several medical imaging applications lack adequately annotated data. Inspired by this, the proposed work leverages unsupervised GAN for skin lesion segmentation. To begin with the methodology, we first briefly discuss generator and discriminator concepts. An adversarial network comprises a generator $(G)$ and a discriminator $(D)$. The generator maps a random vector \textbf{$\gamma$} from source domain space \textbf{$\alpha$} to generate the desired output in the target domain \textbf{$\beta$} and tries to fool the discriminator. $D$ learns to classify whether \textbf{$\beta$} is real (reference ground truth) or fake (generated by $(G)$). The generator’s distribution $p_{G}$ learns over $\alpha$ data, input noise distribution is defined as $P_\gamma$($\gamma$),which maps data space as $(G)$($\gamma$; $\theta_{G}$), where differentiable function $(G)$ has parameters $\theta_{G}$. $(D)$($\alpha$) is the probability distribution of $\alpha$  from the data instead of $p_{G}$.

The adversarial training is represented by following equation \cite{goodfellow} which is minmax game between $G$ and $D$  :

\begin{align}
		\min_{G}\max_{D} V(D,G) & = \begin{aligned}[t]
		& E_{\alpha\sim P_{data}(\alpha)} [logD_{\theta _{D}}(\alpha)]\\
		&+ E_{\gamma \sim P_{\gamma}(\gamma)}[log(1 - logD_{\theta _{D}}(G_{\theta _{G}}(\gamma)))]
		\end{aligned}
\end{align}

where $V$ is function of Discriminator $(D)$, Generator $(G)$,$\gamma$ is from a input noise distribution  $P_{\gamma}(\gamma)$, true samples are from $P_{data}(\alpha)$ and $\theta_{G}$ are generator paramaters and $\theta_{D}$ are discriminator paramaters.

\subsection*{Segmentation Framework}
Generally, segmentation frameworks consist of encoder-decoder-based architecture. The encoder module is the block for feature extraction to capture spatial information within the image. It reduces the spatial size, i.e. the dimension of the input image, and decreases feature map resolution to catch significant level features. The decoder recuperates the spatial data by upsampling the feature map extracted by layers of the encoder and providing the output segmentation map. We propose to modify the architecture design of the encoder-decoder to capture the dense feature map rather than the traditional encoder and change the decoder appropriately, as shown in Figure \ref{seg}. Including squeeze and excitation-based compound scaled encoders significantly improves efficiency in terms of results. 

\subsubsection*{Design of Encoder}

Advancement of CNN designs is dependent on the accessibility of infrastructure and, afterward, the scaling of the model in terms of width $(w)$, depth $(d)$, or resolution $(r)$ of the network to accomplish further significant improvement in performance when there is an expansion in the availability of resources. Instead of doing this scaling manually and arbitrarily, Tan \textit{et al.} \cite{EfficientNet} proposed a novel systematic and automatic scaling approach by introducing a compound coefficient. The novel technique of compound coefficient $\phi$ to efficiently scale the network's depth, width, and resolution with a proper arrangement of scaling factors is per the following equation: 
\vspace{-7mm}
\begin{center}
\begin{equation}
\begin{split}
	w : Network \: width = \beta^{\phi} \\
	d: Network \: depth = \alpha^{\phi}  \\
    r : Input \:Resolution = \gamma^{\phi}  \\
	satisfying \: \alpha \times \beta^{2} \times \gamma^{2} \approx 2 \\
	also \: \alpha\: \geq 1 \: \beta  \geq 1 \: & \: \gamma \geq 1
	\end{split}
\end{equation}
\end{center}

		

The encoder is built using the above equation proposed by Baheti \textit{et al.} \cite{effunet}, consisting of seven building blocks. Each basic building block for this encoder model is squeezing, and excitation functions \cite{8578843} with mobile inverted bottleneck convolution (MBConv), as shown in Figure \ref{seg}(b). Also, swish activation is used in each encoder block, enhancing performance.

\subsubsection*{Design of Decoder}
The encoder downsamples the input image to a smaller resolution and captures contextual information. A decoder block likewise called an upsampling path, comprises many convolutional layers that progressively upsample the feature map obtained from the encoder. The conventional segmentation framework like UNet \cite{Ronneberger} has symmetric encoder and decoder architectures. The proposed architecture builds upon a compound scaled squeeze \& excitation-based encoder and decoder as an asymmetric network. The output features from the encoder are expanded in the decoder blocks consisting of bilinear upsampling. The low-level features from the encoder are combined with the higher-level feature maps from the decoder of respective sizes to generate a more precise segmentation output.

\begin{figure*}[!t]
	\begin{center}
		\includegraphics[scale=0.45]{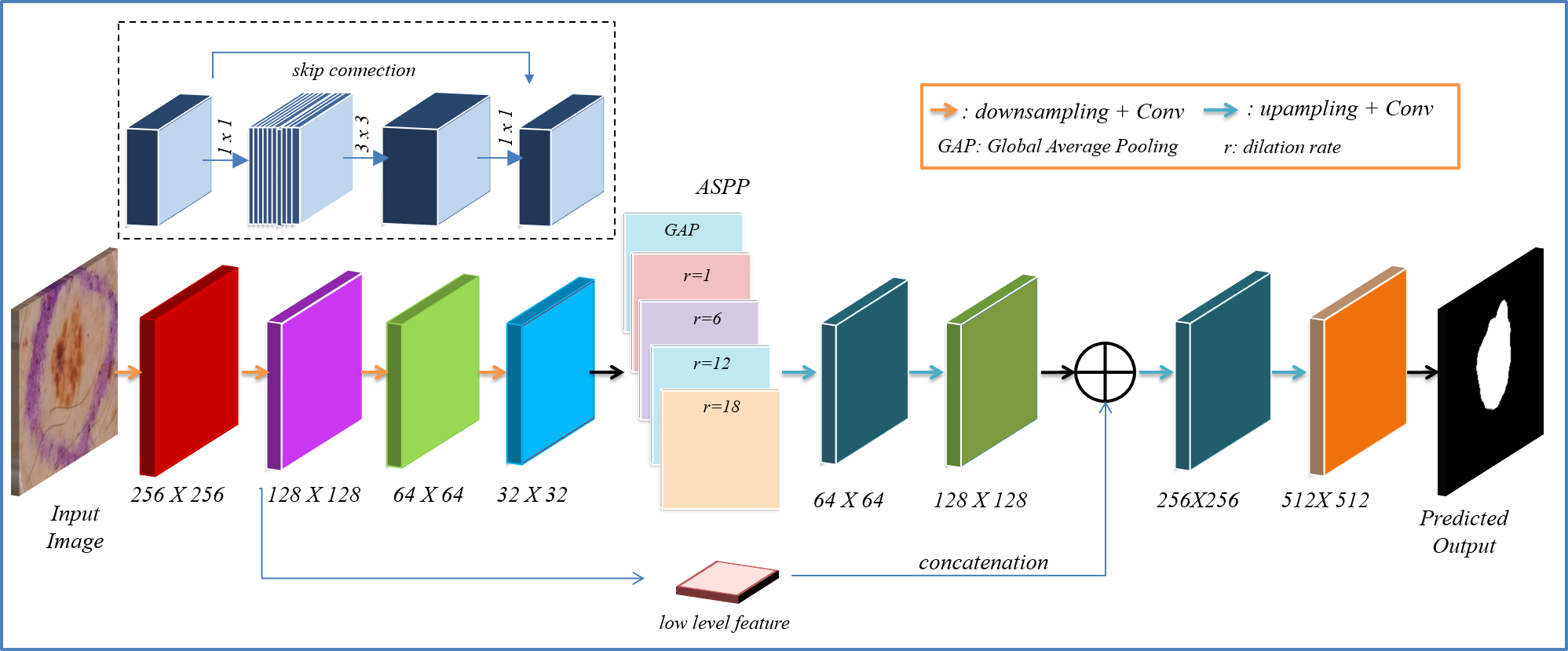}
		\caption{The architecture of the lightweight and efficient segmentation network MGAN. This architecture is based on an inverted residual network and atrous spatial pyramid pooling. The inverted residual block is shown above Encoder.}
		\label{mobile_deeplabv3}
	\end{center}
\end{figure*}

\subsubsection*{Design of lightweight segmentation framework}
To develop a lightweight segmentation architecture for the generator, we leverage the power of MobileNetV2 \cite{MobileNetV2} and DeepLabV3+ \cite{Deeplabv3} consisting of atrous spatial pyramid pooling module (ASPP) as shown in Figure \ref{mobile_deeplabv3}.  MobileNetV2 uses depthwise separable convolution and inverted residual blocks as the basic building module, as shown in Figure \ref{mobile_deeplabv3} above the encoder. MobileNetV2 is modified such that the output stride, i.e., the ratio of the input image to the output image, is 8. It has fewer computations and parameters and is thus suitable for real-time applications. The ASPP block has a variety of dilation rates, i.e., 1, 6, 12, and 18, to generate multi-scale feature maps and further integrate by concatenation. This feature map is upsampled and integrated with a low-level intermediate feature map from the contracting path, i.e., encoder, to generate fine-grained segmentation output. The feature extraction consisted of blocks of inverted residual blocks, as shown in Figure \ref{mobile_deeplabv3}. The stride of the latter blocks is set as one. Images of size 512 $\times$ 512 $\times$ 3 are fed as input to MGAN architecture.

		

\subsection*{Discriminator}
In our architecture, we have a generator and a discriminator. The discriminator supervises the generator to produce precise masks that match the original ground truth. We have implemented a patchGAN-based approach to achieve this, classifying each $m \times n$ mask as equivalent to the ground truth. The discriminator consists of five Conv2D layers with a kernel size of 4 $\times$ 4 and a stride of 2 $\times$ 2, with 64, 128, 256, 512, and 1 feature maps in each layer. LeakyReLU activation with an alpha value of 0.2 is used in each Conv2D layer, with the last layer using sigmoid activation. The patch-based discriminator has an output size ($m \times n$) of 16 $\times$ 16, where one pixel is linked to a patch of input probability maps with a size of 94 $\times$ 94. The discriminator classifies each patch as either fake or real. This learning strategy enforces the predicted label to be similar to the ground truth. The number of parameters is the same as proposed in patchGAN \cite{patchgan}.

We practice the following adversarial technique for each generated label to align with the ground truth labels. A min-max two-step game alternatively renews the generator and discriminator network with adversarial learning. The discriminator function is given by: 
\vspace{-6mm}
\begin{center}
\begin{equation}
	L_{D}(x,y) = -\sum _{x,y} \gamma log(D(I_S)) + (1 - \gamma) log(1 - D(I_T))
\end{equation}
\end{center}

where $x,y$ are the pixel locations of the input,  $D$($I_S$) is the Discriminator function of Source Domain Images($I_S$), i.e., Label Image,  $D$($I_T$) is Discriminator function of Target Domain Images ($I_T$), i.e., Predicted Image and $\gamma$ is the probability of the predicted pixel, $\gamma$ =1 when prediction is from ground truth, i.e., source domain, and $\gamma$= 0 when prediction is from generator segmented mask, i.e., target domain. 

\subsection*{Loss Function} 
We implement smoothing loss based on morphology to improve skin lesions segmentation and supervise the network that captures the lesion's smoothness and fuzzy boundaries. The network's loss function includes dice coefficient loss $(L_{DL})$ as well as the morphology-based smoothing loss $(L_{SL})$. The dice coefficient loss assesses the cross-over between the ground truth and prediction and is given by the condition:
\vspace{-5mm}
\begin{center}
\begin{equation}
	L_{DL(\widehat{v},v))} = 1 - \frac{ 2\sum_{i \in \omega} \widehat{v_{i}}\cdot v_{i}}{\rule{0pt}{0.75em} \sum_{i \in \omega} \widehat{v_{i}}^{2} + {\sum_{i \in \omega} v_{i}^{2}}} 
	\end{equation}
\end{center}

where $\omega$ is the cumulative of pixels in the input image, $v$, and $\widehat{v}$ are the original mask and predicted mask probability map, respectively.

The morphology-based smoothing loss strengthens the network to allow smooth predictions within the nearest neighbor area \cite{8643416}.
It is pairwise interaction of binary labels written as: 
\begin{align*}
L_{SL(\widehat{y},y)} = \sum _{{i \in \Omega}}\sum _{{j \in \mathbb{N^{\iota}}}} B(i,j) \times y_{i} \times \mathopen|\widehat{y}_{i} - \widehat{y}_{j} \mathclose|           &   &where:
	B_{i,j} = \left\{\begin{matrix}1\hspace{0.5cm} if \hspace{0.2cm} y_{i} = y_{j}
	\\ 
	0 \hspace{0.5cm}  otherwise
	\end{matrix}\right.            
\end{align*}

where $\mathbb{N^{\iota}}$ is four neighbor connection of pixels. $y$ and  $\widehat{y}$ denote the ground truth and prediction probability maps, respectively.
The four connected neighbor algorithm-based smoothing loss encourages the surrounding area of pixel $j$ with center pixel $i$ to produce prediction probabilities similar to the original ground-truth class ($B_{i,j} = 1$).

The combined loss function is written as: 
\vspace{-2mm}
\begin{equation}
    L_{\widehat{y},y} =  L_{DL}(\widehat{y},y) + L_{SL}(\widehat{y},y) 
    \end{equation}

Thus, the complete framework works to optimize the loss function by training the network iteratively~\cite{8643416}.


\bibliography{main}



\section*{Acknowledgements}
The authors are grateful to the Center of Excellence, Signal and Image Processing, Shri Guru Gobind Singhji Institute of Engineering and Technology, Vishnupuri, Nanded, Maharashtra, India, for the research resources. Dr. Bakas was supported by grant NCI: U01CA242871 from National Cancer Institute, and Dr. Guntuku was supported by grant NIMHD: R01MD018340 from the National Institutes of Health. The funders had no role in the design and conduct of the study; collection, management, analysis, and interpretation of the data; preparation, review, or approval of the manuscript; and decision to submit the manuscript for publication.

\section*{Author contributions statement}
S.I. and P.D. conducted and analyzed the analyses, and S.I., B.B., and U.B. wrote the main manuscript text. V.P. helped in the preparation of the figures. B.B. and S.C.G. guided the complete work. S.T., B.B., S.B., and S.C.G. reviewed the manuscript.

\section*{Data Availability}
The dataset is available in ISIC archive publicly at \url{https://challenge.isic-archive.com/data/#2018}

\section*{Code Availability}
The source code is available at \url{https://github.com/shubhaminnani/EGAN}.

\section*{Competing interests}
The authors declare no competing interest.

\end{document}